\documentclass[prl,amsmath,amssymb,aps,twocolumn]{revtex4}
\usepackage{graphicx}
\usepackage{feynmp,color}
\usepackage{float}
\usepackage[caption = false]{subfig}

\usepackage{times}

\newcommand{\eq}[1]{(\ref{#1})}

\newcommand{\bea}{\begin{eqnarray}}
\newcommand{\eea}{\end{eqnarray}}
\newcommand{\beq}{\begin{equation}}
\newcommand{\eeq}{\end{equation}}

\newcommand{\rme}{\mathrm{e}}
\newcommand{\rmd}{\mathrm{d}}
\newcommand{\nn}{\nonumber}
\renewcommand{\epsilon}{\varepsilon}
\newcommand{\nott}[1]{}

\graphicspath{{figures/},{}}

\begin{document}

\bibliographystyle{KAY-notitle}

\renewcommand{\omega}{\gamma}

\title{An exact mapping of the  stochastic field theory for Manna sandpiles to interfaces in random media} 
\author{Pierre Le Doussal}
\affiliation{CNRS-Laboratoire de Physique Th\'eorique de l'Ecole Normale
  Sup\'erieure, 24 rue Lhomond, 75005 Paris, France.}
\author{Kay J\"org Wiese}
  \affiliation{CNRS-Laboratoire de Physique Th\'eorique de l'Ecole Normale
  Sup\'erieure, 24 rue Lhomond, 75005 Paris, France.}

\begin{abstract}
We show that the stochastic field   theory  for directed percolation in presence of
an additional conservation law (the C-DP class) can be mapped {\em exactly} to the continuum theory for the depinning of an elastic interface in short-range correlated quenched disorder. On one line of parameters commonly studied, this mapping leads to   the simplest overdamped
dynamics. Away from this line, an additional memory term
arises in the interface dynamics; we argue that
it does not change the universality class. Since C-DP is believed to describe the Manna class
of self-organized criticality, this shows that Manna stochastic sandpiles and disordered elastic interfaces
(i.e.\ the quenched Edwards-Wilkinson model) 
share the same universal large-scale  behavior.
\end{abstract}
\maketitle


Self-organized criticality (SOC) and scale-free avalanches arise
in a variety of models: deterministic and stochastic sandpiles \cite{BakTangWiesenfeld1987,Manna1991,ChristensenCorralFretteFederJossang1996,Dhar1999b,PruessnerBook,Jensen1998},
propagation of  epidemics \cite{JanssenOerdingVanWijlandHilhorst1999},
and 
elastic objects slowly driven in random media
\cite{NarayanDSFisher1993a,NarayanMiddleton1994,DSFisher1998,RossoLeDoussalWiese2009a,LeDoussalWiese2011a,LeDoussalWiese2012a,LeDoussalWieseMoulinetRolley2009}.
In the last decade several authors found evidence that most of
these models  belong to a small number of common universality classes. 
A unifying framework was proposed based on the 
theory
of {\it absorbing phase transitions} (APT) \cite{MarroDickman1999,HenkelHinrichsenLubeck2008}. 
These are non-equilibrium phase transitions, which occur in a vast number of
systems 
between an active 
state and one --or many-- absorbing states. The generic universality class 
in the absence of additional symmetries or conservation laws is the {\it directed-percolation class} (DP) \cite{Grassberger1981,Janssen1981}.
The spreading exponents of the critical DP clusters are interpreted as avalanche 
exponents in the corresponding SOC system \cite{HenkelHinrichsenLubeck2008}. In presence of additional conservation laws,
other classes may arise: An important one is the  {\em conserved directed percolation} class (C-DP),
with an infinite number of
absorbing states. Indeed, it is now accepted, though unproven, that the continuum fluctuation theory for the C-DP class 
provides the effective field theory for the activity in Manna sandpiles \cite{BonachelaMunoz2008}.


Stochastic sandpiles are cellular automata where the toppling rule contains
randomness which is  renewed at each toppling. A notable
example is the Manna model \cite{Manna1991}, a stochastic variant of the deterministic 
Bak-Tang-Wiesenfeld (BTW) sandpile \cite{BakTangWiesenfeld1987}:
{\em Randomly throw grains on a lattice. If the height at one point is $\ge 2$, then move two grains from this site to randomly chosen neighboring sites.} 
This model is not Abelian, i.e.\ the order in which the sites are updated matters.
An Abelian version was proposed by Dhar \cite{Dhar1999b,Dhar1999c}.
Careful numerical studies \cite{Alava2003,BonachelaChateDornicMunoz2007,HuynhPruessnerChew2011,HuynhPruessner2012},
starting with Manna himself \cite{Manna1991}, showed that the Manna model 
and the BTW belong to different universality classes (see \cite{Alava2003,Luebeck2004,PruessnerBook} 
for reviews). Coarse-grained evolution equations for the Manna class were proposed
in \cite{DickmanVespignaniZapperi1998,VespignaniDickmanMunozZapperi1998}. If $\rho(x,t)$ 
denotes the local activity of the sandpile, i.e.\ the local density of unstable sites,
 and $\phi(x,t)$ the local density of grains, 
then they obey the stochastic continuum equations for the C-DP class:
\bea \label{Manna1}
\partial_{t}  \rho(x,t) &=&
a \rho (x,t)-b \rho (x,t)^2+D_{\rho} \nabla^{2} \rho (x,t) \nn\\
&& +\sigma  \eta (x,t) \sqrt{\rho (x,t)}+\omega  \rho (x,t) \phi (x,t)\ , ~~~~~\\
\partial_{t}\phi(x,t) &=& (D_{\phi}\nabla^{2}-m^{2}) \rho(x,t)\ .~~~~~
\label{Manna2}
\eea 
The  parameters $b,D_\rho,D_\phi$ are positive;   
 $\eta(x,t)$ is a (centered) spacio-temporal white noise,
\beq
\left< 
\eta(x,t)\eta(x',t')\right> =\delta^{d}(x-x')\delta(t-t')\ .
\eeq
Clearly, $\rho(x)=0$ with arbitrary ``background'' field $\phi(x,t)$ forms an infinite set of (time-independent) absorbing states.
The  field $\phi(x,t)$ encodes the likeliness of absorbing configurations
to propagate activity when perturbed. From (\ref{Manna2}) $\phi$ is a conserved field 
for $m=0$, reflecting conservation of the total number of grains.
The derivation of (\ref{Manna1})-(\ref{Manna2}) was made more precise in \cite{Pastor-SatorrasVespignani2000,RossiPastor-SatorrasVespignani2000}, 
where it is claimed that all ``stochastic models with an infinite number of absorbing states, in which the order-parameter evolution is coupled to a nondiffusive conserved field, define a unique universality class'', the C-DP. This is further supported in \cite{DickmanMunozVespignaniZapperi2000,MunozGrinsteinDickmanLivi1996}. The C-DP class is believed to contain conserved lattice-gas models, conserved threshold-transfer processes, and others 
\cite{RossiPastor-SatorrasVespignani2000,HenkelHinrichsenLubeck2008,vanWijland2002}.

On the other hand, there were early conjectures that sandpile models and 
disordered elastic manifolds  belong to the same  universality classes: 
The first claim relates the BTW model and 
elastic manifolds driven in a {\it periodic} disorder, i.e.\ charge-density waves \cite{NarayanMiddleton1994},
reexamined recently \cite{FedorenkoLeDoussalWiese2008a}.
It was soon followed by a conjecture \cite{PaczuskiBoettcher1996} on the equivalence of the Oslo 
model \cite{ChristensenCorralFretteFederJossang1996} 
to an elastic string driven by its endpoint in a {\it non-periodic} quenched random field. 
The random field emerges from the stochastic noise in the rule.
Finally, it was conjectured that Manna sandpiles are equivalent to interfaces 
in random media \cite{AlavaLauritsen2001}. These claims are based on exact, 
or almost exact, mappings, onto elastic manifolds with highly discretized evolution rules, 
and it is not  clear what such discretization does to the model
(see e.g. \cite{BonachelaChateDornicMunoz2007} for discussion). 

Quite naturally, it was then proposed that  C-DP and  depinning of an interface in
a quenched random medium belong to the same universality class
\cite{VespignaniDickmanMunozZapperi1998,VespignaniDickmanMunozZapperi2000,DickmanMunozVespignaniZapperi2000,AlavaMunoz2002,BonachelaChateDornicMunoz2007}. 
Until now, however, this remarkable claim is  mainly based on the numerical coincidence of critical exponents
in simulations of discrete models, believed to belong to the respective universality classes 
\cite{BonachelaChateDornicMunoz2007,BonachelaMunoz2008}. 
Ideally, one  wants  a direct connection at the level of the continuum 
theories. The field theory of interfaces subject to disorder is well known,
  described by the functional RG (FRG) method,  involving an infinite number (a function)
of relevant couplings  near its upper critical
dimension $d_{\rm c}=4$ \cite{DSFisher1986}. It describes depinning 
\cite{NattermannStepanowTangLeschhorn1992,ChauveLeDoussal2001,LeDoussalWieseChauve2002} 
and, more recently, avalanches \cite{LeDoussalWiese2008c,LeDoussalWiese2011a,LeDoussalWiese2012a}. Hence one would like to relate it to the field theory of the C-DP class. Although it was
realized that  renormalization  of the C-DP class is more complex than that of 
 standard DP  which requires only a few couplings, 
the attempts to handle it were unsuccessful \cite{vanWijland2002,Wijland2003}. More intriguingly, 
the full renormalized disorder correlator was measured numerically \cite{BonachelaAlavaMunoz2008}, 
and found indistinguishable from that of random interfaces obtained
in \cite{RossoLeDoussalWiese2006a}.

The aim of this Letter is to provide an exact mapping 
in the continuum, between the C-DP 
class defined by Eqs.~(\ref{Manna1}) and (\ref{Manna2}), and 
an interface
driven in quenched disorder, with a specific, exponentially decaying, microscopic disorder correlator. Along 
a line in parameter space it maps C-DP to the simplest overdamped dynamics of the interface, thereby
providing the long-sought  proof of equivalence of the two systems. Away from this
line, the dynamics of the interface is more complex, and involves a memory kernel.
As we show, it nevertheless falls into the same universality class
as the simplest overdamped model, i.e.\  quenched Edward-Wilkinson (QEW). 

Let us start by considering the two coupled equations of motion  (\ref{Manna1}) and (\ref{Manna2}).
For convenience we  added a parameter $m^{2}$, since it appears in the interface model   as an infrared regulator. Although we are interested in the limit $m\to 0$, it is useful to define the theory with $m>0$, since this insures that the activity $\rho(x,t)$ will stop, even without grains leaving the system, which therefore can be taken infinitely large.
To simplify the identification,  note that by rescaling of space we can set $D_{\rho}\to 1$. By rescaling $\phi(x,t)$, we can then set $D_{\phi}\to 1$.
Finally rescaling both $\rho(x,t)$ and $\phi(x,t)$, we can set $\sigma\to 1$. 
This simplifies the model to 
 \bea \label{Manna1b}
\partial_{t}  \rho(x,t) &=&
a \rho (x,t)-b \rho (x,t)^2+ \nabla^{2} \rho (x,t) \nn\\
&& +  \eta (x,t) \sqrt{\rho (x,t)}+\omega  \rho (x,t) \phi (x,t) ~~~~~\\
\partial_{t}\phi(x,t) &=& (\nabla^{2}-m^{2}) \rho(x,t)\ .~~~~~
\label{Manna2b}
\eea 
The activity variable $\rho(x,t) \geq 0$ for all times 
\cite{endnote57}.
Note that the case $\omega=0$, with $b>0$, corresponds to the field theory of
directed percolation: In the absence of noise, i.e.\ in mean field, it exhibits a 
transition between $\rho>0$ for $a>0$ and $\rho=0$ for $a \leq 0$.
This transition exists in any $d$.  The noise $\eta(x,t)$ becomes relevant for $d \leq d_{\rm c}=4$,
a property shared by DP and  C-DP; the latter has   $\omega>0$ 
which we now examine. 

As we will see below, the case $\omega=b$ is special. We therefore set $b:= \omega+ \kappa$.
We  define new  variables, a {\em force} ${\cal F}(x,t)$ and a {\em velocity} $\dot u(x,t)$ (denoting $\partial_t$ or a dot   time derivatives):
\bea \label{change} 
 {\cal F}(x,t) &:=& \rho(x,t) -\phi(x,t)  -  \frac{a+m^{2}}{\omega} \ , \\
 \rho(x,t) &:=& \dot u(x,t)\ .
\eea
The total number of topplings at position $x$ since $t=0$ is  
 $u(x,t)-u(x,t=0)=\int_0^t \rmd t' \rho(x,t)$. The identification of $u$ as a height 
for the associated elastic interface is standard \cite{BonachelaAlavaMunoz2008},  while the identification of ${\cal F}$ as a  ``force" 
is new.  Clearly, the initial
value of the field $u(x,t=0)$ does not carry any information for the C-DP problem, while it
does for the interface problem
\cite{endnote58}. For notational simplicity we will set  $u(x,t=0)=0$. 
All our results can be extended to the general case by  replacing
$u(x,t) \to u(x,t)-u(x,t=0)$.
The equations of motion for ${\cal F}(x,t)$ and $\dot u(x,t)$  then are
\bea \label{9a}
\partial_{t}{\cal F}(x,t) &=& -\omega {\cal F}(x,t) \dot u(x,t) - \kappa \dot u(x,t)^2  \nn\\
&& + \eta(x,t) \sqrt{\dot u(x,t)}\ ,~~~~ \\
\label{10a}
\partial_{t} \dot u(x,t) &=& [\nabla^{2}-m^{2}] \dot u (x,t) + \partial_t {\cal F}(x,t)\ .~~~~~
\eea
The problem is defined with initial data $\dot u(x,t=0)$ and ${\cal F}(x,t=0)$. 
The second equation (\ref{10a}) can be integrated into 
\bea \label{eqf1} 
\partial_{t} u(x,t) &=& [\nabla^{2}-m^{2}] u (x,t) + {\cal F}(x,t) + f(x) \ ,
\\
 \label{force} 
f(x)&=&\dot u(x,0) - {\cal F}(x,0) =  \phi(x,0)  + \frac{a+m^{2}}{\omega} \ .\qquad
\eea
Eq.~(\ref{eqf1}) describes the 
motion of an elastic interface submitted to a known time-independent external force
$f(x)$, and a space-time dependent force ${\cal F}(x,t)$.
Because of the term $m^2$,  the interface  also sees a quadratic well. 
Integration of Eq.\ (\ref{Manna2b}) shows that the change in the 
background field, $\phi(x,t) - \phi(x,0)$, can be interpreted as the sum of the elastic force
plus the force from the quadratic well, acting on the interface.
Eq.\ (\ref{9a}) determines ${\cal F}(x,t)$ {\it as a functional} of the field $u(x,t)$. 
It is a stochastic functional,  depending on the noise, and is
formally written as ${\cal F}(x,t) \equiv {\cal F}[u , \eta](x,t)$. Once ${\cal F}(x,t)$ is known, substituting it into Eq.~(\ref{eqf1}) defines a problem of an elastic manifold in a
random medium. 
As we show now,  ${\cal F}(x,t)$ can be written {\it explicitly}. 
Eq.\ (\ref{9a}) is linear 
in ${\cal F}$ with two source terms, hence its solution is 
\beq \label{tot} 
 {\cal F}(x,t) =  e^{- \omega u(x,t)} {\cal F}(x,t=0)   +  {\cal F}_{\rm dis}(x,t) + {\cal F}_{\rm ret}(x,t) \ .
\eeq
The first term depends on the initial condition, and decays to zero
if the interface moves by more than $1/\omega$; it can thus be 
ignored in the steady state. 
As we now show, the second term can be interpreted as a quenched random pinning force.
It arises from the noise  in Eq.\ (\ref{9a}), and is independent of $\kappa$. It is 
the only term  when $\kappa=0$ (then ${\cal F}_{\rm ret}=0$) 
i.e.\ for $\omega=b$. 
This term can be written as
$ {\cal F}_{\rm dis}(x,t) = F\big(u(x,t),x\big)$,
where for each $x$, $F(u,x)$ is an Orstein-Uhlenbeck process \cite{VanKampenBook},  solution of
the stochastic equation
\beq \label{OU} 
\partial_u F(u,x) = - \omega F(u,x) + \tilde \eta(x,u) \ ,
\eeq 
with initial data $F(0,x)=0$, and $\tilde \eta(x,u)$  a white noise, uncorrelated in $x$ and $u$. 
A pedestrian way to derive Eq.~(\ref{OU}) is to write the white noise $\eta(x,t)=\rmd B_x(t)/\rmd t$
in Eq.\ (\ref{9a}) in terms of independent one-sided Brownians  $B_x(t)$ indexed by $x$, with $B_x(0)=0$,
and integrate
the linear equation as
\bea
&& \!\!\!{\cal F}_{\rm dis}(x,t)  =  \int_0^t \rmd t' \frac{\rmd B_x(t')}{\rmd t'}  \sqrt{\dot u(x,t')} e^{- \omega [ u(x,t) - u(x,t')]}  \nn \\
& &=  e^{- \omega u(x,t)}  \int_0^{u(x,t)} e^{\omega u} \rmd \tilde B_x(u)  = F\big(u(x,t),x\big)\ .~~~~ \label{manip}
\eea
The force $F(u,x)$ is the solution of the Orstein-Uhlenbeck process (\ref{OU})
in terms of the white noises $\tilde \eta(x,u)=\rmd\tilde B_x(u)/\rmd x$. 
It can be written as a (time-changed) Brownian,
\beq \label{finalOU} 
 F(u,x) = \frac{e^{- \omega u}}{\sqrt{2 \omega}} \tilde B_x\!\left(e^{2 \omega u}-1\right).
\eeq 
The second line in (\ref{manip}) is obtained by noting that under a time change 
$\rmd u=\dot u(x,t) \rmd t$ each Brownian $B_x(t)$ is changed into  another   Brownian $\tilde B_x(u)$ with $\tilde B_x(0)=0$, as
$
 \sqrt{\dot u(x,t')} \rmd B_x(t') = \rmd\tilde B_x\big(u(x,t')\big).
$
Eq.~(\ref{finalOU}) 
is obtained using the identity $\int_0^v f(u) \rmd B_x(u) = \tilde B_x(\int_0^v f(u)^2 \rmd u)$ 
for test functions $f(u)$, 
resulting from the scale invariance of Brownian motion.

Hence, neglecting the first (decaying) term in Eq.\ (\ref{tot}), we  showed 
that along the line $\omega=b$ the C-DP system maps onto
\beq \label{eqinter} 
\partial_{t} u(x,t) = [\nabla^{2}-m^{2}] u (x,t) + F(u(x,t),x) + f(x) \ .
\eeq
This is an interface driven in a quenched random force field $F(u,x)$.
This random field is Gaussian,   specified by its correlator, which can be calculated
from (\ref{finalOU}), 
using $\overline{B_x(u) B_{x'}(u')} = \delta(x-x') \min(u,u')$. As expected, one finds
that the Orstein-Uhlenbeck process
becomes stationary when the interface has been driven on
distances larger than $1/\omega$:
\bea
\label{13b}
 \overline {F(u,x)F(u',x')}&=& \delta^{d}(x-x')\frac{\rme^{- \omega |u-u'|}- \rme^{- \omega (u+u')}}{2 \omega} \nn  \\
&\to&_{\omega u, \omega u' \gg 1} \delta^{d}(x-x') \Delta_0(u-u')~~~~~
\eea
with $\overline{F(u,x)}=0$. The {\em bare disorder correlator} of the
random pinning force thus is
\bea \label{13c} 
\Delta_0(u) = \frac{\rme^{- \omega |u|}}{2 \omega}\ .
\eea 
It is clearly  short-ranged, and as   a peculiarity has  a linear cusp. Usually one considers smooth
microscopic disorder, i.e.\ an analytic $\Delta_0(u)$, 
which under RG (i.e.\ coarse graining)  develops 
a cusp linked to the existence of many metastable states
and avalanches beyond the Larkin scale $L_{\rm c} \sim 1/m_{\rm c}$
\cite{endnote59}.
A cusp in the microscopic disorder means that there
are avalanches of arbitrarily small sizes. On the other hand, it is known from the universality of the interface problem that any short-ranged force-force correlator 
flows at large scale, under coarse-graining, to the same renormalized disorder correlator,  the  universal {\em depinning} fixed point
\cite{endnote59}. 
Its upper critical dimension is $d_{\rm c}=4$, implying that C-DP  also has
$d_{\rm c}=4$. The fixed-point function has been calculated analytically in an $\epsilon=d_{\rm c}-d$ expansion \cite{LeDoussalWieseChauve2002} and measured numerically \cite{RossoLeDoussalWiese2006a}.
It determines the two independent exponents of the depinning transition, the roughness exponent 
$\zeta$ of the field $u \sim L^\zeta$, $\zeta>0$ for $d<d_{\rm c}$, and
the dynamic exponent $z$, $t \sim L^z$, $z<2$ for $d<d_{\rm c}$, and their $\epsilon$-expansions \cite{LeDoussalWieseChauve2002}.

Let us now discuss the correspondence between the active-absorbing phase transitions
for 
C-DP and depinning. For simplicity 
consider a spatially uniform
initial condition $\phi(x,t=0)=\phi$, 
s.t.\ the initial driving force acting on the 
interface in Eq.\ (\ref{force}) is uniform, $f(x)=f$. We now set the control parameter $m\to 0$ so that there is a globally active phase 
corresponding to an interface moving at constant steady-state mean velocity $\overline{\dot u(x,t)}=v  \sim (f-f_{\rm c})^\beta>0$, if 
$f>f_{\rm c}$. Here $f_{\rm c}$ is the depinning threshold force, which is, at least in principle, calculable once
the correlator $\Delta_0$ is known. Translating to the C-DP system it implies an active phase
with $\rho>0$, when $a + \omega \phi  > \omega f_{\rm c}$, and a phase transition where $\rho$ vanishes
with the same exponent $\beta$ as a function of the distance to criticality.   Due to a  symmetry of the interface problem, 
 $\beta=\nu (z-\zeta)=\frac{z-\zeta}{2-\zeta}$.
 By scaling this gives $\rho = \dot u \sim t^{-\theta}$ at criticality with $\theta=1- \frac{\zeta}{z}$, e.g.\ as  response to a (large) specially uniform perturbation at $t=0^{+}$, in the limit of $v\to 0^{+}$. In the language of APT 
 \cite{HenkelHinrichsenLubeck2008} this is a {\em steady-state exponent}.
 
%
 

Let us now consider the protocol for avalanches in the
absorbing phase, near criticality. In the sandpile model (e.g.\ in 
numerical simulations of the Manna model) one usually starts from an initial condition with 
a non-vanishing activity $\rho(x,0)=\dot u(x,0)\ge 0$, either by adding a  single grain, or adding grains in an extended region. This generates an
avalanche which stops when  $\rho(x,t)=0$ for all $x$. For the elastic manifold it
is equivalent to having the interface at rest up to time $t=0$, and then to increase the force by $\dot u(x,0)$.
This procedure is then repeated until one reaches the steady state (for $u(x,t) \gg 1/\omega$), where the  avalanche statistics becomes stationary. 
It is  known for  interfaces that under this procedure 
the system reaches the {\em Middleton attractor}, a sequence of
well-characterized metastable states between successive avalanches \cite{Middleton1992}. Avalanches with this statistics
have well-defined exponents, which were discussed e.g. in \cite{LeDoussalWiese2012a,DobrinevskiLeDoussalWiese2014a,DobrinevskiPhD}.

To summarize, along the line $\omega=b$, i.e.\ $\kappa=0$, we presented an exact and  direct mapping 
(valid in any dimension) between the continuum C-DP  Eqs.~(\ref{Manna1})-(\ref{Manna2}) and 
the simplest model of a driven interface with overdamped dynamics, 
subject to a {\em quenched} random force $F(u(x,t),x)$ with  (microscopic) correlations given by Eq.~\eq{13c}, and confined in a parabolic well of curvature $m^{2}$. This confirms, and makes  precise, the beautiful numerical study  of Ref.\ \cite{BonachelaAlavaMunoz2008}; there the authors observe  that Manna sandpiles, the Oslo model, C-DP as given by Eqs.\ (\ref{Manna1})-(\ref{Manna2}),   and disordered elastic manifolds have the same renormalized (effective) disorder correlator. If one accepts that the Manna class 
coincides with  C-DP, it establishes the long sought exact mapping to disordered elastic manifolds
\cite{endnote60}.
This agreement is valid in the stationary state,  
but our exact mapping establishes the complete correspondence,  and allows to study the evolution from any given initial state.

Some remarks are in order.  The interface equation (\ref{eqinter}) with the choice of correlator \eq{13b} 
possesses a special Markovian property, which it inherits from the force evolution equation (\ref{9a})
(for $\kappa=0$), and which allows it to be
solved without storing the full random-force landscape. 
The latter is constructed as the avalanche proceeds, hence
 is determined only for $u\le u(x,t)$. 
This property was noted in
\cite{DobrinevskiPhD,DobrinevskiLeDoussalWieseprep} 
and can be used for efficient numerics \cite{DobrinevskiPhD,KoltonLeDoussalWieseInPreparation}. 

The limit $\omega \to 0$ is also of interest. If one keeps $\kappa=0$, i.e.\ $b \to 0$, one sees from (\ref{force}) and (\ref{tot}) that in that limit
\beq \label{bfm} 
\dot u(x,t) - \dot u(x,t=0) = [\nabla^{2}-m^{2}] u (x,t) + \tilde B_x\big(u(x,t)\big) 
\eeq
This is the so-called Brownian force model (BFM), the mean-field
theory for  avalanches of the interface model \cite{LeDoussalWiese2012a,DobrinevskiLeDoussalWiese2011b,LeDoussalWiese2011b}.
 If we keep $b>0$,  the limit instead is towards the DP class.

Let us finally discuss the C-DP system for $\kappa \neq 0$, i.e.\ away from the special line
$\omega=b$ in Eq.~(\ref{Manna1b}). If the new source term 
$\kappa \dot u^2$, which appears in Eq.~(\ref{9a}) for $\partial_t {\cal F}$, were 
directly inserted into Eq.~(\ref{10a}) for $\dot u$, 
the mapping to the interface would fail, as such a term is  relevant \cite{endnote61}.
Fortunately, this term is {\it screened} by the short-range disorder,
and instead of being relevant  is only marginal. To show this, let
 us come back to Eq.\ (\ref{tot}), which now has a second contribution,
\beq
{\cal F}_{\rm ret}(x,t)  = - \kappa \int_0^t \rmd t'\, \dot u(x,t')^2\, e^{- \omega [ u(x,t) - u(x,t')]} \ .
\eeq
This term can be rewritten using integration by parts as 
\bea 
{\cal F}_{\rm ret}(x,t)  &=& \frac{ \kappa}{\omega}e^{- \omega u(x,t) } \left[ \dot u(x,0)+ \int_0^t   \rmd t'\, \ddot u(x,t')\, e^{ \omega u(x,t')} \right] \nn\\
&& - \frac{ \kappa}{\omega} \dot u(x,t)\ .
\eea
Inserting into Eq.~(\ref{eqf1}) we  finally obtain the equation of motion
\bea
\frac{b}{\omega} \partial_{t} u(x,t) &=& [\nabla^{2}-m^{2}] u (x,t) + F(u(x,t),x) + f(x)  \nn \\
&& +  \frac{\kappa}{\omega}  \int_0^t \rmd t' \ddot u(x,t') e^{- \omega [ u(x,t) - u(x,t')]} \nn \\
 && +  \left[ \frac{b}{\omega}  \dot u(x,0)  - f(x) \right] e^{- \omega u(x,t)}  \label{final} \ .
\eea
We recall $f(x) = \phi(x,0)+\frac{a+m^2}{\omega}$. This equation is
 equivalent to 
  the C-DP system (\ref{Manna1})-(\ref{Manna2})
for $\rho(x,t)=\dot u(x,t)$ with specified initial data $\dot u(x,0)$, $\phi(x,0)$,
and is a salient result of our letter. Note that it results from a simple
change of variables, which maps a system with {\it annealed noise}, the
C-DP, to a system with {\it quenched noise}, the interface; as such it bears some analogy to the Cole-Hopf transformation 
used to solve the Kardar-Parisi-Zhang (KPZ) equation. 

Let us now discuss Eq.~(\ref{final}). 
The first line  describes the standard overdamped equation of motion 
of the interface, with the same random force $F(u,x)$ as before, but a new friction
coefficient $b/\gamma$. The third line depends on the initial condition and decays to
zero when the interface has moved by more than $1/\omega$. As we
now focus on the stationary regime we can  neglect it. The second line is a new 
memory term.
To estimate its relevance at large scales, 
consider the large-$\gamma$ limit, and
replace $e^{- \gamma z} \to \frac{1}{\gamma} \delta(z)$, hence
$e^{- \gamma [ u(x,t) - u(x,t')]}  \to \frac{1}{\gamma \dot u(x,t)} \delta(t-t')$. 
The second line of (\ref{final}) then becomes
\bea \label{corr1}  
\frac{\kappa}{\gamma^2} \partial_t \ln \dot u(x,t) + {\cal O}({\gamma^{-3}})
\eea 
where each power of $1/\omega$ in the expansion yields terms which are  more and more
irrelevant by power counting, since each power of $1/\gamma$ 
comes with a power of $1/u \sim L^{-\zeta}$. This argument indicates 
that the new term is marginally irrelevant, and only shifts the numerical
value of the friction. Hence we conclude that the universality class 
of C-DPand of the QEW model should be the same, even for $b \neq \omega$. 

The present work calls for further studies: First, Eq.~(\ref{final})
can be analyzed  using FRG to confirm our
conclusions and  explore 
this
unusual interface dynamics. Our work opens the way to 
study, within a common RG framework, a variety of models ranging 
from interfaces to absorbing phase transitions. It can be extended to long-range 
elasticity (long-range toppling),
or to a variety of perturbations. The simplest one is to add $m^2 \dot w(x,t)$ to
each of the Eqs.~(\ref{Manna1})-(\ref{Manna2}) in order to reproduce   the standard
driving for the interface \cite{LeDoussalWiese2011a}. Another extension is 
the crossover to DP as
both $\omega$ and $b$ are small.

Second, Eq.~(\ref{final})
 permits to study initial conditions, hence
to disentangle effects dominated by transients from
those of  the Middleton attractor. 
That  allows to  treat  avalanches with localized
seeds in the context of  APTs, used to define  
{\em spreading exponents}. E.g.\ the survival probability in C-DP, $P_{\mbox{\scriptsize C-DP}}^{\rm surv}(t)\sim t^{-\delta}$ is related to the avalanche-duration distribution  at depinning, $P_{\rm dep}(T)\sim T^{-\alpha}$, via $\delta = \alpha-1 =(d-2+\zeta)/z$. We checked that indeed $\delta =0.17$ and $0.48$ in $d=1$ and $2$, both for depinning, see table 2 of \cite{DobrinevskiLeDoussalWiese2014a}, and Manna sandpiles \cite{BonachelaMunoz2008,BonachelaPhD}.

%


Third, since our mapping is local in space, it can be extended to finite-size systems with
 prescribed boundary conditions, in order to study the case $m=0$. 
 Imposing $\rho(x,t)=\phi(x,t)=0$ at the boundary  corresponds
 to the common choice to let grains ``fall off'' from the boundary. In our variables  it implies $u(x,t)=\dot {\cal F}(x,t)=0$ at the boundary.
 
 Finally one should go back to Ref.\ \cite{BonachelaAlavaMunoz2008},  and understand cusps
in a more general setting. 
A  challenging questions is whether the quenched KPZ class
can be treated in a similar setting,  especially since  
some of its exponents in $d=1$ are described by DP. Other unsolved problems, 
such as DP with quenched disorder \cite{HenkelHinrichsenLubeck2008} may now
be studied. 

In conclusion, we have provided an {\it exact mapping} from the field theory of 
a reaction-diffusion system with an additional conservation law, the C-DP system
of Eqs.~(\ref{Manna1})-(\ref{Manna2}), to a specific continuum model of 
an interface driven in a random landscape. Using universality 
we show that the C-DP class, Manna stochastic sandpiles 
and the quenched Edwards-Wilkinson model belong to a single, and hence
{\it very large} universality class which spans self-organized criticality,
avalanches in disordered systems, and reaction-diffusion models. This raises 
the prospect of a unified field theory for all these systems using functional RG methods. It also defines a framework, in which probabilists could put this claim
on {\it rigorous} grounds, 
as was recently done for the KPZ class
\cite{hairer,corwin}. 

We thank A. Dobrinevski for very useful discussions
and acknowledge support from PSL grant ANR-10-IDEX-0001-02-PSL.
We thank KITP for hospitality and support in part by the NSF 
under Grant No. NSF PHY11-25915.


\begin{thebibliography}{10}
\bibitem{endnote58}{The arbitrariness in this choice leads to a symmetry, the statistical tilt symmetry (STS), which constrains the exponents. The elasticity is not renormalized and $\nu =1/(2-\zeta )$. }
\bibitem{endnote61}{$\kappa \protect \mathaccentV {dot}05Fu^2 \sim L^{2 \zeta -2 z}$ is more relevant than $\nabla ^2 \protect \mathaccentV {dot}05Fu \sim L^{\zeta -z-2}$ at large scales, since $\zeta +2-z >0$ for $d<d_{\protect \rm  c}$.}
\end{thebibliography}

%

\end{document}